# Cell-Laden alginate biomaterial modelling using three-dimensional (3D) microscale finite element technique

Abhinaba Banerjee
Department of Mechanical Engineering,
Indian Institute of Engineering Science and Technology,
Shibpur, Howrah- 711103, West Bengal, India

Sudipto Datta
Centre for Healthcare Science and Technology,
Indian Institute of Engineering Science and Technology,
Shibpur, Howrah- 711103, West Bengal, India

Amit Roy Chowdhury
Department of Aerospace Engineering and Applied Mechanics,
Indian Institute of Engineering Science and Technology,
Shibpur, Howrah- 711103, West Bengal, India

Pallab Datta
Department of Pharmaceutics, National Institute of Pharmaceutical Education and Research,
Kolkata 700054, West Bengal, India

A novel modelling technique using finite element analysis to mimic the mechanoresponse of cell-laden biomaterial is proposed for the use in bioinks and other tissue engineering applications. Here a hydrogel-based composite biomaterial specimen was used consisting of 5% (V/V) HeLa cells added to alginate solution (4% W/V) and another specimen with no living cell present in alginate solution (4% W/V). Tensile test experiments were performed on both the specimens with a load cell of 25 N. The specimens were bioprinted using an in-house developed three-dimensional (3D) bioprinter. To allow for the nonlinear hyperelastic behavior of the specimen, the specimens were loaded very slowly, at rates of 0.1 mm/min and 0.5 mm/min, during the tensile test. The microscale finite element models developed in Ansys were loaded with similar load rates and their responses were recorded. Both the model results were validated with the experiment results. A very good agreement between the finite element model and the tensile test experiment was observed under the same mechanical stimuli. Hence, the study reveals that bioprinted scaffold can be virtually modeled to obtain its mechanical characteristics beforehand.

## I. INTRODUCTION

The last few decades witnessed a huge progress in the field of biomaterial and tissue engineering in terms of developing novel biomaterials and producing patient-specific bioscaffolds. This [1] review paper highlights 3D printing of skin tissue which enables the construction of biomaterials and living cells in the desired pattern. Although considerable success has been achieved in the printing of cells and biomaterials, the road to 3D printing of complete functional organs is still largely untrodden due to organ-level complexity. The labor, cost, and time-intensive in-vitro experiments add further to this complexity. Only a few years ago researchers at Harvard Wyss institute were able to 3D print a vascularized miniature heart ventricle [2]. In recent years, it is seen that huge computational power has been harnessed to develop 3D models from computed tomography (CT) data, which mimics the mechanical properties of actual biomaterial in the experiment, with great accuracy [3], [4]. Another commonly used method to model and predict biomaterial response is the numerical technique. The three-dimension bio-printability prediction of fibrillar structure bioinks using numerical methods has been recently done in [5].

The most commonly used materials for scaffold-based soft biomaterials are fibrin, agarose, chitosan, hyaluronic acid, alginate sulfate, Matrigel-TM, Poly (methyl methacrylate) (PMMA), and collagen. These materials exhibit excellent mechanical properties and are biocompatible. Researchers have found that these biomaterials experience high shear stresses during 3D printing and consideration of these forces is important to develop biomaterials with a high degree of functionality [6]. All the finite element techniques available in the literature to model biomaterials is done using macroscale models which require lesser computational resource. But these models become inaccurate when biomaterials are cell-laden, which incorporates material property variation at the microscale level. Microscale-based simulation techniques are computation resource and time-intensive. But they can predict the biomaterial response to external stimuli to a great degree of accuracy. No research paper in the literature has been reported that uses microscale finite element analysis to model biomaterials incorporated with living cells.

The present study aims to model Alginate biomaterial infused with living HeLa cells using microscale finite element technique. The biomaterial specimen taken is in the form of a single strand that has been bioprinted using an in-house 3D bioprinter. The specimen is then tensile tested experimentally. The nonlinear hyperelastic property of alginate has been obtained from the experiment and the material property of the HeLa cell is obtained from the already available literature. The random distribution of HeLa cells within the alginate strand volume is achieved by writing a code in Python software. The computational model was also tensile tested with similar loading as that of the experiment.

## II. MATERIALS AND METHODS

Sodium salt alginic acid was procured from Sigma-Aldrich USA, HeLa cell from National Centre of Cell Science (NCCS), Pune in India, and ASTM Type III grade water from Wasser Lab, Navarra, Spain. For conducting the experiments two specimens were bioprinted. Both the specimen had 4% (W/V) alginate. One specimen does not contain any HeLa cell and the other specimen has a 5% (V/V) Hela cell in alginate. The live HeLa cell was added with the alginate solution using a micropipette in a petri dish. Then the mixture was stirred for 24 hours using a microtip to ensure uniform volumetric distribution of the Hela cell in the alginate solution. The specimen details are summarized in Table 1. Both the printed specimens are cylindrical in shape with a diameter of 500 μm and length of 60 mm.

**TABLE I: STOICHIOMETRIC COMPOSITION OF THE SPECIMENS**

|  | Sodium Alginate salt (g) | ASTM Type III Grade water (ml) | HeLa cell per 5 ml of alginate solution (μl) 25 million cells are present in 1 ml solution |
|---|---|---|---|
| Specimen 1 | 4 | 100 | 0 |
| Specimen 2 | 4 | 100 | 250 |

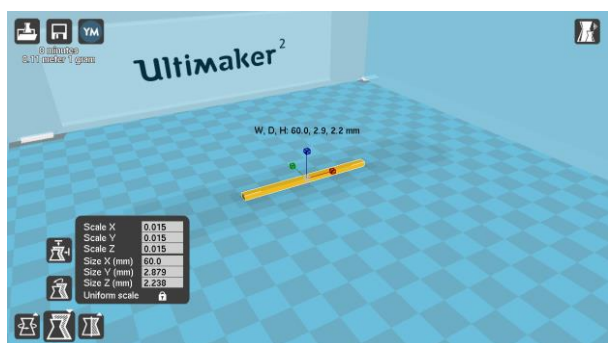

**Figure 1: CURA model of the strand.**

Bioprinting the strand specimen requires prior 3D modelling of the software, which is done in Sketchup software (Trimble Inc., Sunnyvale, California) and CURA-TM version 15.04.4 (shown in Fig.1), and finally uploading it into the 3D printer. The bioprinter used to print our model is a custom-built machine allowing X, Y, and Z movements. Provision for orthogonal compensation and auto bed-leveling is also incorporated into the machine.

To ensure the accuracy of the printing, it was done at speed of 50 mm/s using a 0.5 mm diameter nozzle. The temperature of the alginate-HeLa cell ink was maintained at 250 C. Post printing the strands were cross-linked using 1mL of 0.5 Molar CaCl2 solution. Finally, the specimens were dried in an oven at 400 C. The dried specimens were checked for their dimensional accuracy under a microscope (Nikon ECLIPSE Ti) attached to a Nikon Digital camera. The diameters that were obtained were 502.91 μm and 504.57 μm (Fig. 2).

## III. MATERIAL CHARACTERIZATION AND MATERIAL PROPERTY DETERMINATION

To perform the tensile tests on the specimen a UTS machine of Tinus Olsen, UK make was used. The specimens were clamped with 15 mm of length inside the sample holder, leaving a gauge length of 30 mm. Ten preconditioning cycles with 10% strain were performed prior to the actual tension test to check the consistency. Experimental tensile testing setup shown in Fig. 3. 'Horizon' software was used to collect the tensile test data. Before using the raw data for further analysis, it was smoothened in 'ORIGIN' software using the LOWESS technique.

The material modelling of hyperelastic materials is done using the strain energy function, which eventually requires the determination of the coefficients of the strain energy equation. To obtain these material coefficients we checked the tension test data with the already available material models in ABAQUS software. The reduced polynomial, N=2 fitted the experimental data best and was hence selected as the material model for alginate. HeLa cell material is assumed to be isotropic in nature with Poisson's ratio equals 0.38 and young's modulus equals 0.45 as found in the literature[7], [8].

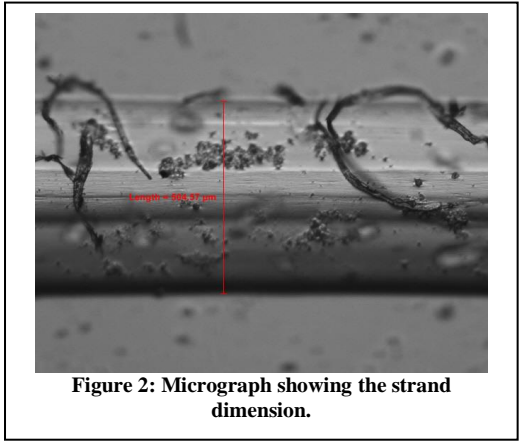

**Figure 2: Micrograph showing the strand dimension.**

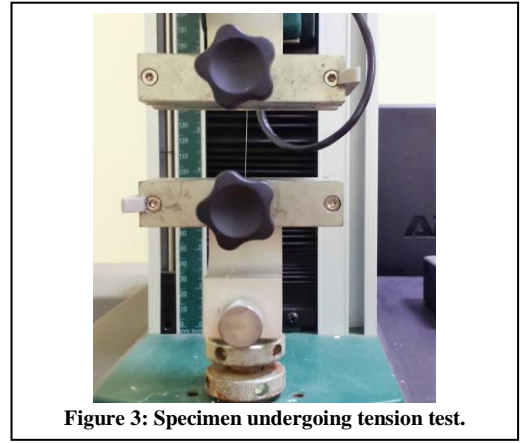

**Figure 3: Specimen undergoing tension test.**

## IV. FINITE ELEMENT MODELLING

The three-dimensional (3D) modelling of the strand was done in SolidWorks software (version 2018) with a diameter of 0.5 mm and length of 30 mm. This model was then exported to ANSYS APDL (version 2019) for meshing and further analysis. The 3D cylindrical model has been meshed with the 10-node tetrahedral element and 90 μm mesh size. HeLa cell has an average diameter of 30 μm and we have assumed three HeLa cells aggregate together wherever present inside the solid volume. To mimic the random distribution of the cells within the volume we wrote a code in Python which manipulates the ANSYS element list file incorporating the variation of material. One end of the cylinder was constricted of all DOF and the load was applied on the other end. The cylindrical model developed with the initial boundary conditions is shown Fig. 4. The large displacement key was turned on during analysis to include the hyperelasticity effects of alginate.

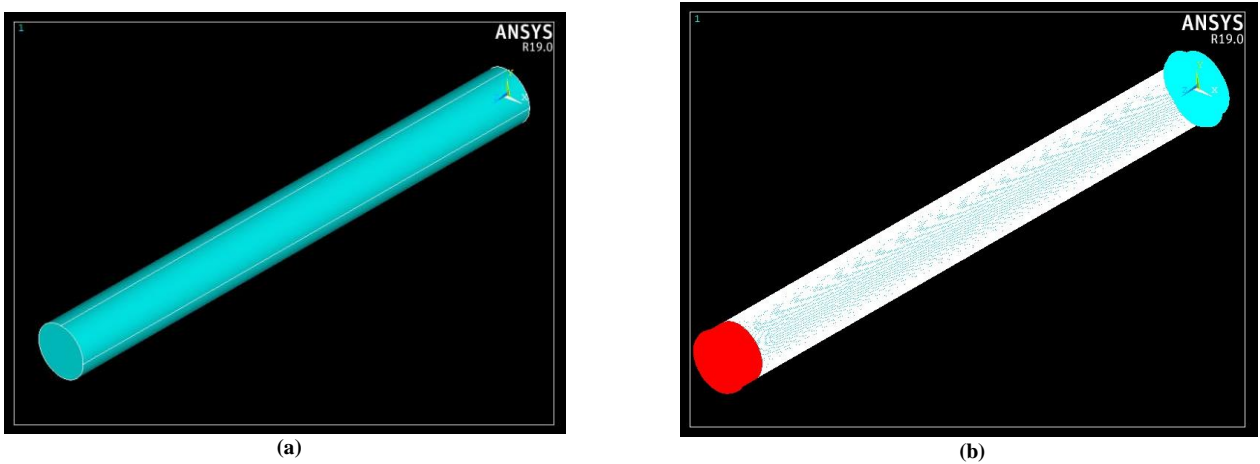

**Figure 4: Illustration of (a) strand model generated in ANSYS, (b) Load and boundary constraints applied to the model.**

## V. RESULTS AND DISCUSSION

*A. Experimental Tension Test*

Initially, the specimen with no HeLa cell was put under tension test. The ultimate load at which the cylindrical model failed was found to be 0.145 N, applied at a rate of 0.1 mm/min on the model. The elongation in the model at maximum load condition was 16.3%. The ultimate stress at which the material failed was 0.722 MPa, as found in the 'Horizon' software. This test was followed by tensile testing the specimen with a 5% (V/V) HeLa cell. Here, the specimen could endure a load of 0.175 N just before failure. The corresponding elongation in the specimen was found to be 21.57%. The normal stress in the specimen as calculated in 'Horizon' software was 0.889 MPa.

*B. Finite Element Model*

This section highlights the results obtained from the FE analysis when loading conditions of the experiment were copied into the 3D ANSYS model. The maximum load which could be applied for the model without any HeLa cell was 0.142 N. This load was applied to the model in steps to capture the nonlinear material properties of alginate. The corresponding Von-Mises stress in the model was 0.72 MPa. The elongation was found to be 0.161 mm/mm at maximum load. Fig. 5 depicts the stress versus strain comparison for the experiment and the FE analysis for only alginate model. When HeLa cells are absent in the specimen our model predicts both the Von-Mises stress and strain with great accuracy as depicted in Fig. 5.

The second FE analysis was done on alginate-HeLa composite model. Here, the maximum load that could be applied before failure was found to be 0.142 N. The Von-Mises stress in the specimen was calculated to be 0.74 MPa at maximum loading condition. The corresponding elongation in the 3D model was 0.175 mm/mm. Fig. 6 shows the comparison of stress versus strain between the experiment tensile test and the FE model analysis. It is seen from the graph that the finite element model developed is able to replicate the non-linear behavior of the hyperelastic alginate material.

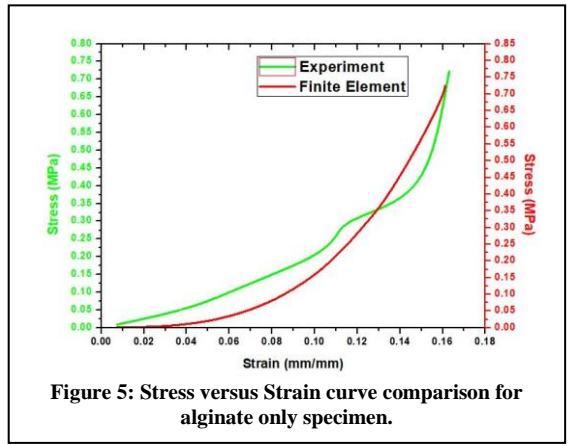

**Figure 5: Stress versus Strain curve comparison for alginate only specimen.**

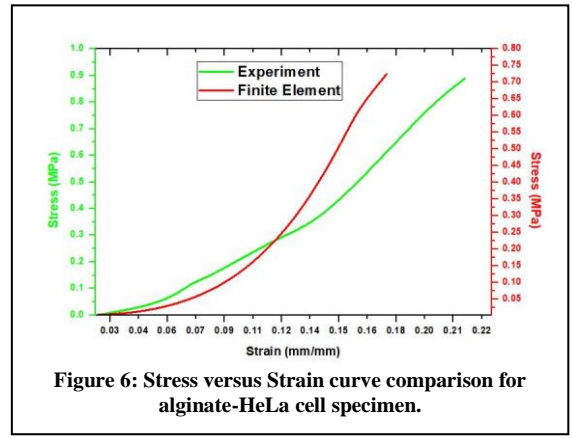

**Figure 6: Stress versus Strain curve comparison for alginate-HeLa cell specimen.**

## V. CONCLUSION

In the present paper, a novel microscale finite element modelling technique is proposed to mimic biomaterial mechanical properties under a tensile test. The model also captures the microscale mechanical property variation at the cell biomaterial interface. The stress versus strain comparison restates that this modelling technique can be used to replace actual experiments which require time, effort, and money. These modelling techniques can be further extended to model other biomaterials and also scaffold structures of much more complexity. The only limitation of mimicking complex structures using microscale modelling is the requirement of huge computational resources. Generally, for tissue engineering applications the cell seeding concentration in the biomaterial is restricted to low values (~ 5% V/V). These cells proliferate with time in the biomaterial matrix. Such variation of cell concentration can be captured using four-dimensional (4D) printing and modelling techniques.